\pdfoutput=1
\documentclass{article}
\usepackage{etoolbox}

\usepackage[nonatbib,final]{neurips_arxiv}
\usepackage{natbib}

\newtoggle{release}
\toggletrue{release}

\usepackage{natbib}
\usepackage[utf8]{inputenc}
\usepackage[T1]{fontenc}
\usepackage{amsfonts}       
\usepackage{amsmath, amssymb, amsthm}
\usepackage{booktabs}       
\usepackage{dsfont}
\usepackage{enumerate}
\usepackage{graphicx}
\usepackage{hyperref}
\usepackage{microtype}      
\usepackage{nicefrac}       
\usepackage{pgfplots}
\usepackage{pgf}
\usepackage{tikzscale}
\usepackage{tikz}
\usepackage{url}
\usepackage{xcolor}
\usepackage{subcaption}
\usepackage{ifthen}
\usepackage{pifont}
\usepackage{makecell}
\usepackage{scalerel}
  
\usepackage{pgfplotstable}
\pgfplotsset{compat=newest}
\usepgfplotslibrary{external}
\usetikzlibrary{external,arrows,calc,positioning,shapes.geometric,math}

\nottoggle{release}{
\newcommand{\nico}[1]{{\color{blue}#1}}
\newcommand{\alex}[1]{{\color{purple}#1}}
\newcommand{\leon}[1]{{\color{green}#1}}
\newcommand{\francis}[1]{{\color{orange}#1}}
}{
\newcommand{\nico}[1]{}
\newcommand{\neil}[1]{}
\newcommand{\alex}[1]{}
\newcommand{\leon}[1]{}
\newcommand{\francis}[1]{}
}

\DeclareMathOperator{\Conv}{Conv1d}

\DeclareMathOperator{\Convtr}{ConvTr1d}
\DeclareMathOperator{\ReLU}{Relu}
\DeclareMathOperator{\GLU}{GLU}
\DeclareMathOperator{\SDR}{SDR}

\DeclareMathOperator{\SIR}{SIR}
\DeclareMathOperator{\SAR}{SAR}

\providecommand{\norm}[1]{\left\|#1\right\|}

\providecommand{\reel}{\mathbb{R}}

\DeclareMathOperator*{\bigadd}{\scalerel*{+}{\sum}}

\newcommand{\chmark}{\ding{51}}
\newcommand{\crmark}{\ding{55}}
\newcommand{\source}[1]{\texttt{#1}}

\title{Music Source Separation in the Waveform Domain}
\author{
  Alexandre D\'efossez \\
  Facebook AI Research \\
  INRIA \\
  \'Ecole Normale Sup\'erieure\\
  PSL Research University\\
  \texttt{defossez@fb.com}
  \And
  Nicolas Usunier\\
  Facebook AI Research \\
  \texttt{usunier@fb.com}
  \And
  L\'eon Bottou\\
  Facebook AI Research \\
  \texttt{leonb@fb.com}
  \And
  Francis Bach\\
  INRIA \\
  \'Ecole Normale Sup\'erieure\\
  PSL Research University\\
  \texttt{francis.bach@ens.fr}
}

\definecolor{mydarkblue}{rgb}{0,0.08,0.45}
\hypersetup{ %
    pdftitle={Music Source Separation in the Waveform Domain},
    pdfauthor={D\'efossez, Usunier, Bottou, Bach},
    pdfsubject={Source separation},
    pdfkeywords={deep learning, source separation, audio processing, waveform},
    pdfborder=0 0 0,
    pdfpagemode=UseNone,
    colorlinks=true,
    linkcolor=mydarkblue,
    citecolor=mydarkblue,
    filecolor=mydarkblue,
    urlcolor=mydarkblue,
    pdfview=FitH}

\begin{document}

\maketitle

\begin{abstract}
 Source separation for music is the task of isolating contributions, or \emph{stems}, from different instruments
    recorded individually and arranged together to form a song. Such components include voice, bass, drums and any other accompaniments.
Contrarily to many audio synthesis tasks where the best performances are achieved by models that directly generate the waveform, the state-of-the-art in source separation for music is to compute masks on the magnitude spectrum. 
In this paper, we compare two waveform domain architectures. We first adapt Conv-Tasnet, initially developed for speech source separation,
to the task of music source separation. While Conv-Tasnet beats many existing spectrogram-domain methods, it suffers
from significant artifacts, as shown by human evaluations. We propose instead Demucs, a novel waveform-to-waveform model,
with a U-Net structure and bidirectional LSTM.
Experiments on the MusDB dataset show that, with proper data augmentation, Demucs beats all
existing state-of-the-art architectures, including Conv-Tasnet, with 6.3 SDR on average, (and up to 6.8 with 150 extra training songs, even surpassing the IRM oracle for the bass source).
Using recent development in model quantization, Demucs can be compressed down to 120MB
without any loss of accuracy.
We also provide human evaluations, showing that Demucs benefit from a large advantage
in terms of the naturalness of the audio. However, it suffers from some bleeding,
especially between the \source{vocals} and \source{other} source.
\end{abstract}

\section{Introduction}

Cherry first noticed the ``cocktail party effect''~\citep{cocktail}: how the human brain is able to
separate a single conversation out of a surrounding noise
from a room full of people chatting.
Bregman later tried to understand how the brain was
able to analyse a complex auditory signal and segment it into higher level streams.
His framework for auditory scene analysis~\citep{asa} spawned its
computational counterpart, trying to reproduce or model accomplishments
of the brains with algorithmic means~\citep{casa}, in particular regarding source separation capabilities.

When producing music, recordings of individual instruments called \emph{stems}
are arranged together and mastered into the final song. The goal of source
separation is to recover those individual stems from the mixed signal.
Unlike the cocktail party problem, there is not a single source of interest to
differentiate from an unrelated background noise, but instead a wide variety
of tones and timbres playing in a coordinated way. In the SiSec Mus
evaluation campaign for music separation~\citep{sisec},
those individual stems were grouped into 4 broad categories: (1)~\source{drums}, (2)~\source{bass}, (3)~\source{other}, (4)~\source{vocals}.
Given a music track which is a mixture of these four sources, also called the mix, the goal is to generate four waveforms that correspond to each of the original sources. We consider here the case of supervised source separation, where the training data contain music tracks (i.e., mixtures), together with the ground truth waveform for each of the sources.

State-of-the-art approaches in music source separation still operate on the spectrograms generated by the short-time Fourier transform (STFT). They produce a mask on the magnitude spectrums for each frame and each source, and the output audio is generated by running an inverse STFT on the masked spectrograms reusing the input mixture phase
\citep{sony_densenet,sony_denselstm,takahashi2020d3net}. Several architectures trained end-to-end to directly synthesize the waveforms have been proposed \citep{wavenet_sep,waveunet_singing},
but their performances are far below the state-of-the-art: in the last SiSec Mus evaluation campaign~\citep{sisec}, the best model that directly predicts waveforms achieves an average signal-to-noise ratio (SDR) over all four sources of $3.2$, against $5.3$ for the best approach that predicts spectrograms masks (also see Table~\ref{table:comparison} in Section \ref{sec:results}).
An upper bound on the performance of all methods relying on masking spectrograms is given by the SDR obtained when using a mask computed
using the ground truth sources spectrograms,
for instance the Ideal Ratio Mask (IRM) or the Ideal Binary Mask (IBM) oracles.
For speech source separation,
\citet{convtasnet} proposed Conv-Tasnet, a model that reuses the masking approach of spectrogram methods but learns the masks jointly with a convolutional front-end,
operating directly in the waveform domain for both the inputs and outputs. Conv-Tasnet surpasses both the IRM and IBM oracles.

\begin{figure}
    \centering
    \includegraphics[width=0.33\textwidth,trim={0 0 0 2cm},clip]{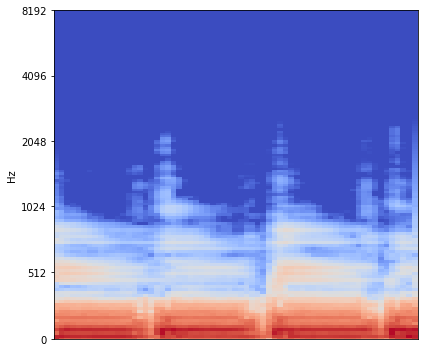}%
    \includegraphics[width=0.33\textwidth,trim={0 0 0 2cm},clip]{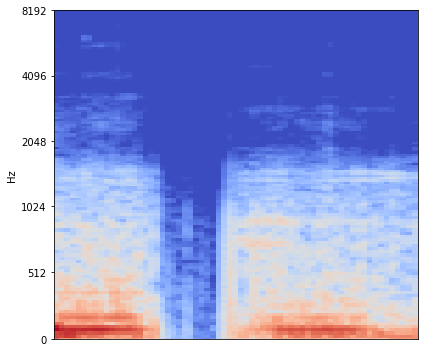}%
    \includegraphics[width=0.33\textwidth,trim={0 0 0 2cm},clip]{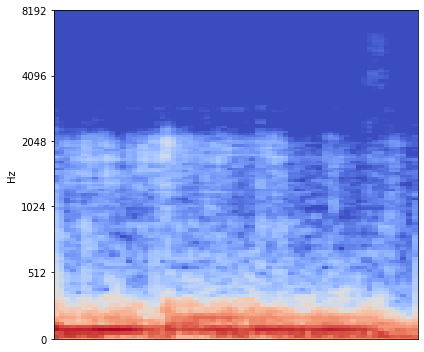}
    \caption{Mel-spectrogram for a 0.8 seconds extract of the \source{bass} source from the track ``Stich Up'' of the MusDB test.
    From left to right: ground truth, Conv-Tasnet estimate and Demucs estimate. We observe that Conv-Tasnet missed one note entirely.}
    \label{fig:mel}
\end{figure}

Our first contribution is to adapt the Conv-Tasnet architecture, originally designed for monophonic speech separation and audio sampled at 8 kHz, to the task of stereophonic music source separation for audio sampled at 44.1 kHz.
While Conv-Tasnet separates with a high accuracy the different sources, we observed artifacts when listening to the generated audio: a constant broadband noise, 
hollow instruments attacks or even missing parts. They are especially noticeable on the \source{drums} and \source{bass} sources and we give one example on Figure~\ref{fig:mel}, and measure it with human evaluation in Section~\ref{sec:results}.


To overcome the limitations of Conv-Tasnet, we introduce Demucs, a novel architecture for music source separation. Similarly to Conv-Tasnet, Demucs is a deep learning model that directly operates on the raw input waveform and generates a waveform for each source. Demucs is inspired by models for music synthesis rather than masking approaches. It is a U-net architecture with a convolutional encoder and a decoder based on wide transposed convolutions with large strides inspired by recent work on music synthesis \citep{sing}. The other critical features of the approach are a bidirectional LSTM between the encoder and the decoder,
increasing the number of channels exponentially with depth,
gated linear units as activation function \citep{glu} which also allow for masking, and a new initialization scheme.

We present experiments on the MusDB~\citep{musdb} benchmark. While Conv-Tasnet outperforms several existing spectrogram-domain methods,
it does suffer from large audio artifacts as measured by human evaluations.
On the other hand, with proper augmentation, the Demucs architecture surpasses
all existing spectrogram or waveform domain architectures in terms of SDR~\citep{measures}, with 6.3 points of SDR without extra training data (against 6.0 for the best existing method D3Net), and up to 6.8 with extra training data. In fact, for the \source{bass} source, Demucs
is the first model to surpass the IRM oracle, with 7.6 SDR (against 7.1 for the IRM). We confirm the usefulness of pitch/tempo shift
augmentation, as first noted by ~\citet{cohen2019improving}, with a gain of 0.4 points of SDR, in particular for a model with a large number of parameters like Demucs, while it can be detrimental to Conv-TasNet.

We discuss in more detail the related work in the next Section. We then describe the original Conv-Tasnet model of \citet{tasnet} and its adaptation to music source separation. Our Demucs architecture is detailed in Section \ref{sec:model}. We present the experimental protocol in Section \ref{sec:exp}, and the experimental results compared to the state-of-the-art in Section \ref{sec:results}. Finally, we describe the results of the human evaluation and the ablation study.

\section{Related Work}

A first category of methods for supervised music source separation work
on time-frequency representations. They predict a power spectrogram for each source
and reuse the phase from the input mixture to synthesise individual waveforms.
Traditional methods have mostly focused on blind (unsupervised) source separation.
Non-negative matrix factorization techniques~\citep{nmf_unified} model the power spectrum
as a weighted sum of a learnt spectral dictionary, whose elements are
grouped into individual sources.
Independent component analysis~\citep{ica} relies on independence assumptions
and multiple microphones to separate the sources.
Learning a soft/binary mask over power spectrograms has been done using
either HMM-based prediction~\citep{one_mic_hmm} or segmentation techniques~\citep{bach2005blind}.

With the development of deep learning, fully supervised methods have gained momentum.
Initial work was performed on speech source separation~\citep{classif_dnn_separation}, followed by works on
music using simple fully connected networks over few spectrogram frames~\citep{uhlich2015deep},
LSTMs~\citep{uhlich2017improving}, or multi scale convolutional/recurrent networks~\citep{liu2018denoising,sony_densenet}.
\citet{multichannel_deep} showed that Wiener filtering is an efficient post-processing step for spectrogram-based models
and it is now used by all top performing models in this category.
Those methods have performed the best
during the last SiSec 2018 evaluation~\citep{sisec} for source separation on the MusDB~\citep{musdb} dataset.
After the evaluation, a reproducible baseline called Open Unmix has been released by \citet{openunmix} and matches the top submissions trained only on MusDB.
 MMDenseLSTM, a model proposed by~\citet{sony_denselstm}
and trained on 807 unreleased songs currently holds the absolute record of SDR in the SiSec campaign. 
More recently, \citet{takahashi2020d3net} improved the spectrogram-domain state-of-the-art with D3Net, which uses
dilated convolutions with dense connection. \citet{spleeter2020} propose another U-Net architecture
for spectrogram masking, called Spleeter, that is trained on the Bean dataset~\citet{pretet2019singing}, composed of short extracts from nearly 25,000 songs. While Spleeter provides strong performance and has
been widely adopted by the digital music industry, it is now outperformed by more advanced spectrogram domain architectures like D3Net, even though it is trained on much more data.

More recently, models operating in the waveform domain have been developed, so far with
worse performance than those operating in the spectrogram domain.
A convolutional network with a U-Net
structure called Wave-U-Net was used first on spectrograms~\citep{waveunet_singing} and then
adapted to the waveform domain~\citep{wavunet}. Wave-U-Net was submitted to the SiSec 2018 evaluation campaign
with a performance inferior to that of most spectrogram domain models by a large margin. 
A Wavenet-inspired, although using a regression loss and not auto-regressive, was first used
for speech denoising~\citep{wavenet_denoising} and then adapted to source
separation~\citep{wavenet_sep}. Our model significantly outperforms Wave-U-Net.
Given that the Wavenet inspired model performed worse than Wave-U-Net, we did not consider it for comparison.

In the field of monophonic speech source separation, 
spectrogram masking methods have enjoyed good performance~\citep{speech_fourier,speech}.
\citet{tasnet} developed Tasnet, a waveform domain methods using masking over a learnable front-end
obtained from a LSTM
that reached the same accuracy. Improvements 
were obtained by \citet{unfolding} for spectrogram methods using the unfolding of a few iterations of a phase 
reconstruction algorithm in the training loss. In the mean time, \citet{convtasnet}
refined their approach, replacing the LSTM with a superposition of dilated convolutions, which improved the 
SDR and definitely surpassed spectrogram based approaches, including oracles that use the ground truth sources such as the ideal ratio mask (IRM) or the ideal binary mask (IBM). 
We show that, while surpassing many existing spectrogram domain methods, Conv-Tasnet does not benefit much from pitch/tempo shift
augmentation, and we could only reach a SDR of 5.7 with it. Besides, its audio samples suffer from significant audio artifacts as measured
by human evaluations. However, it has less contamination from other sources than Demucs.
\citep{luo2020dual} introduced the Dual-Path RNN architecture for speech separation. \citet{nachmani2020voice} improved on it,
with strong results on the MusDB dataset (5.8 SDR). It is unclear whether it could benefit from pitch/tempo shift augmentation.

\citet{samuel2020meta} also adapted the Conv-Tasnet architecture to the task of music source separation, using meta-learning (i.e., 
the weights for each source are generated from a meta neural network), multi-resolution separation and auxilary losses. Their model matches
the performance of our Conv-Tasnet baseline, but would likely suffer from the same limitation.
Finally, \citet{lancaster2020frugal} trained an original Tasnet model, with 30 times more training data than in the Musdb dataset. While this model achieves strong results, it fails to be competitive when considering with models like Conv-Tasnet or Demucs, especially considering the large amount of training data used.

\section{Adapting Conv-Tasnet for music source separation}
\label{sec:tasnet}
We describe in this section the Conv-Tasnet architecture of \citet{tasnet} and give the details of how we adapted the architecture to fit the setting of the MusDB dataset.

\paragraph{Overall framework}
Each source $s$ is represented by a waveform $x_s \in \reel^{C, T}$
where $C$ is the number of channels (1 for mono, 2 for stereo) and $T$
the number of samples of the waveform. The mixture (i.e., music track) is the sum of all sources $x := \sum_{s=1}^S x_s$.  We aim at training a model $g$ parameterized by $\theta$, such that $g(x) = (g_s(x; \theta))_{s=1}^S$, where $g_s(x; \theta)$ is the predicted waveform for source $s$ given $x$, 
that minimizes
\begin{equation}
    \label{eq:separation}
    \min_\theta \sum_{x\in\mathcal{D}} \sum_{s=1}^S L(g_s(x; \theta), x_s)
\end{equation}
for some dataset $\mathcal{D}$ and reconstruction error $L$. The original Conv-Tasnet
was trained using a loss called scale-invariant source-to-noise ratio (SI-SNR), similar to the SDR loss described in Section~\ref{sec:exp}.
We instead use a simple L1 loss between the estimated and ground truth sources. We discuss in more details regression losses in the context of our Demucs architecture in Section \ref{sec:loss}.

\paragraph{The original Conv-Tasnet architecture }
Conv-Tasnet \citep{tasnet} is composed of a learnt front-end 
that transforms back and forth between the input monophonic mixture waveform sampled at 8 kHz and a 128 channels over-complete representation sampled at 1 kHz using a convolution as the encoder and a transposed convolution as the decoder, both with a kernel size of 16 and stride of 8.
The high dimensional representation is masked through a separation network composed of stacked residual blocks. Each block is composed of a 
a 1x1 convolution, a PReLU~\citep{prelu} non linearity, a layer wise normalization over all channels jointly~\citep{ba2016layer}, a depth-wise separable convolution ~\citep{chollet2017xception,howard2017mobilenets} with a kernel size of 3, a stride of 1 and a dilation of $2^{n \,\text{mod}\, N}$, with $n$ the 0-based index of the block and $N$ an hyper-parameter, and another PReLU and normalization. The output of each block participates to the final mask estimation through a skip connection, preceded by a 1x1 convolution.
The original Conv-Tasnet counted $3 \times N$ blocks with $N = 8$.
The mask is obtained summing the output of all blocks
and then applying ReLU. The output of the encoder is multiplied by the mask and before going through the decoder.

\paragraph{Conv-Tasnet for music source separation}
We adapted their architecture to the task of stereophonic music source separation: the original Conv-Tasnet has a receptive field of 1.5 seconds for audio sampled at 8 kHz, we take $N = 10$ and increased the kernel size (resp. stride) of the encoder/decoder from 16 (resp. 8) to 20 (resp. 10), leading to the same receptive field at 44.1 kHz. We observed better results using $4 \times N$ blocks instead of $3 \times N$ and $256$ channels for the encoder/decoder instead of 128.
With those changes, Conv-Tasnet obtained
state-of-the-art performance on the MusDB dataset, surpassing all known spectrogram based methods by a large margin as shown in Section~\ref{sec:results}.


\paragraph{Separating entire songs}
The original Conv-Tasnet model was designed for short sentences of a few seconds at most. When evaluating it on an entire track, we obtained the 
best performance by first splitting the input track into chunks of 8 seconds each. We believe this is because of the global layer normalization. During training, only small audio extracts are given, so that a quiet part or a loud part would be scaled back to an average volume. However, when using entire songs as input, it will most likely
contain both quiet and loud parts. The normalization will not map both to the same volume, leading to a difference between training and evaluation.
We did not observe any side effects when going from one chunk to the next, so we did not look into fancier overlap-add methods.
\section{The Demucs Architecture}
\label{sec:model}

\begin{figure}
\vskip -9mm
  \centering
  \def\pscale{0.65}
  \begin{subfigure}[b]{0.49\textwidth}
    \centering
    \begin{tikzpicture}[
    every node/.style={scale=\pscale},
    conv/.style={shape=trapezium,
        trapezium angle=70, draw, inner xsep=0pt, inner ysep=0pt,
        draw=black!90,fill=gray!5},
    deconv/.style={shape=trapezium,
        trapezium angle=-70, draw, inner xsep=0pt, inner ysep=0pt,
        draw=black!90,fill=gray!5},
    linear/.style={draw, inner xsep=1pt, inner ysep=1pt,
        draw=black!90,fill=green!5},
    rnn/.style={rounded corners=1pt,rectangle,draw=black!90,
        fill=blue!5,minimum width=0.6cm, minimum height=0.6cm},
    skip/.style={line width=0.2mm, ->},
]
    \def\yshift{0.3em}
    \def\base{9cm}
    \def\dec{0.55cm}

    \node (base) at (0, 0) {};
    \def\sourcea{0.25 * 2 * (0.5 * x - floor(0.5 * x))}
    \def\sourceb{0.25 * exp(-(x + 10)/10) * cos(deg(4 * x))}
    \def\sourcec{0.25 * cos(deg(0.3 * x)) * cos(deg(2 * x + 0.9 * cos(deg(4 * x)))}
    \def\sourced{0.25 * cos(deg(5 * x) * (
        1 + 0.1 * cos(deg(1 * x)))
     )}
    \begin{axis}[
        anchor=north,
        at=(base),
        scale=0.6,
        domain=-10:10,
        axis y line=none,
        axis x line=none,
        samples=200,
        color=black,
        height=2.5cm,
        width=\base + 3cm]
          \addplot[mark=none] {
              (
                \sourcea + \sourceb + \sourcec + \sourced
               )
         };
    \end{axis}
    \node (e1) [conv, minimum width=\base - \dec, anchor=south] at (0, 0)
        {$\mathrm{Encoder}_1(C_{in}=2, C_{out}=64)$};
    \node (e2) [conv, minimum width=\base - 2*\dec, anchor=south] at
        ($(e1.north) + (0, \yshift)$)
        {$\mathrm{Encoder}_2(C_{in}=100, C_{out}=128)$};
    \node (edots) [conv, minimum width=\base - 3*\dec, anchor=south] at
        ($(e2.north) + (0, \yshift)$)
        {$\ldots$};
    \node (e6) [conv, minimum width=\base - 4*\dec, anchor=south] at
        ($(edots.north) + (0, \yshift)$)
        {$\mathrm{Encoder}_6(C_{in}=1600, C_{out}=2048)$};

    \node (ls0) [rnn] at ($(e6.north) + (-0.35 * \base + 2 * \dec,0.4cm)$) {L};
    \foreach \k/\text in {1/S,2/T,3/M} {
        \tikzmath{
            int \prev;
            \prev=\k - 1;
        }
        \node (ls\k) [rnn,anchor=west] at ($(ls\prev.east) + (0.4cm, 0)$) {\text};
        \draw [<->] (ls\prev) -- (ls\k);
    }
    \node (ls3) [anchor=west] at ($(ls3.east) + (0.1cm, 0)$) [align=left] {hidden size=2048\\2 bidirectional layers};

    \node (linear) [linear, minimum width=\base - 5*\dec, anchor=south]
        at ($(e6.north) + (0, 0.8cm)$) {$\mathrm{Linear}(C_{in}=4096, C_{out}=2048)$};
    \node (d6) [deconv, minimum width=\base - 4*\dec, anchor=south] at
        ($(linear.north) + (0, \yshift)$) {$\mathrm{Decoder}_6(C_{in}=2048, C_{out}=1024)$};
    \node (ddots) [deconv, minimum width=\base - 3*\dec, anchor=south] at
        ($(d6.north) + (0, \yshift)$) {$\ldots$};
    \node (d2) [deconv, minimum width=\base - 2*\dec, anchor=south] at
        ($(ddots.north) + (0, \yshift)$) {$\mathrm{Decoder}_2(C_{in}=128, C_{out}=64)$};
    \node (d1) [deconv, minimum width=\base - \dec, anchor=south] at
        ($(d2.north) + (0, \yshift)$) {$\mathrm{Decoder}_1(C_{in}=64, C_{out}=4 * 2)$};

    \path[skip] (e1.west) edge[bend left=45] node [right] {} (d1.west);
    \path[skip] (e2.west) edge[bend left=45] node [right] {} (d2.west);
    \path[skip] (edots.west) edge[bend left=45] node [right] {} (ddots.west);
    \path[skip] (e6.west) edge[bend left=45] node [right] {} (d6.west);
    \newcommand\myoutput[3]{
        \begin{axis}[
            anchor=south,
            scale=0.6,
            at=#1,
            domain=-20:20,
            axis y line=none,
            axis x line=none,
            samples=200,
            height=2.5cm,
            color=#2,
            width=\base + 3cm]
            \addplot[mark=none] {
                #3
            };
        \end{axis}
    }
    \node (o1) at (d1.north) {};
    \node (o2) at ($(o1.north) + (0, 4mm)$) {};
    \node (o3) at ($(o2.north) + (0, 4mm)$) {};
    \node (o4) at ($(o3.north) + (0, 4mm)$) {};
    \myoutput{(o1)}{violet}{\sourcea}
    \myoutput{(o2)}{olive}{\sourceb}
    \myoutput{(o3)}{red}{\sourcec}
    \myoutput{(o4)}{blue}{\sourced}
\end{tikzpicture}
    \caption{Demucs architecture with the mixture waveform as input
    and the four sources estimates as output. Arrows represents U-Net
    connections.\label{fig:fullmodel}}
  \end{subfigure}%
  \hfill%
  \begin{subfigure}[b]{0.49\textwidth}
    \centering
    \begin{tikzpicture}[
    every node/.style={scale=\pscale},
    conv/.style={shape=trapezium,
        trapezium angle=70, draw, inner xsep=0pt, inner ysep=2pt,
        draw=black!90,fill=gray!5},
    deconv/.style={shape=trapezium,
        trapezium angle=-70, draw, inner xsep=0pt, inner ysep=2pt,
        draw=black!90,fill=gray!5},
    rewrite/.style={shape=rectangle,
        draw, inner xsep=11pt, inner ysep=3pt,
        draw=black!90,fill=gray!5},
    inout/.style={rounded corners=1pt,rectangle,draw=black!90,
        fill=violet!5,minimum width=0.6cm, minimum height=0.6cm},
    skip/.style={line width=0.2mm, ->},
    sum/.style      = {draw, circle, fill=gray!5, inner xsep=0pt, inner ysep=0pt},
]
    \def\yshift{0.3em}
    \def\base{8cm}
    \def\dec{0.55cm}
    \def\deltax{1cm}

    \node (base) at (0, 0cm) {};
    \node (pad) at (0cm, -1.5cm) {};
    \node (conv) [conv, minimum width=\base - \dec, anchor=south] at
      (base.north) {$\GLU(\Conv(C_{in}, 2 C_{in}, K=3, S=1))$};
    \node (deconv) [deconv, minimum width=\base - \dec, anchor=south] at
      ($(conv.north) + (0,\yshift)$) {$\ReLU(\Convtr(C_{in},C_{out}, K=8, S=4))$};

    \def\yshift{0.6em}
    \node (skip) [inout, anchor=north] at ($(conv.south) - (\deltax, 3 * \yshift)$) {$\mathrm{Encoder}_i$};
    \node (sum) [sum] at ($(conv.south) - (0, 1.5 * \yshift)$) {$\bigadd$};

    \draw[->]  (skip.north) -- ($(conv.south) - (\deltax, 1.5 * \yshift)$) -- (sum.west);

    \node (prev) [inout, anchor=north] at ($(conv.south) - (-\deltax, 3 * \yshift)$) {$\mathrm{Decoder}_{i+1}$ or LSTM};
    \draw[->]  (prev.north) -- ($(conv.south) - (-\deltax, 1.5 * \yshift)$) -- (sum.east);

    \draw[->]  (sum.north) -- (conv.south);

    \node (next) [inout, anchor=south] at ($(deconv.north) + (0, \yshift)$) {$\mathrm{Decoder}_{i-1}$ or output};
    \draw[->]  (deconv.north) -- (next.south);
\end{tikzpicture}
    \begin{tikzpicture}[
    every node/.style={scale=\pscale},
    conv/.style={shape=trapezium,
        trapezium angle=70, draw, inner xsep=0pt, inner ysep=2pt,
        draw=black!90,fill=gray!5},
    deconv/.style={shape=trapezium,
        trapezium angle=-70, draw, inner xsep=0pt, inner ysep=2pt,
        draw=black!90,fill=gray!5},
    rewrite/.style={shape=rectangle,
        draw, inner xsep=8pt, inner ysep=3pt,
        draw=black!90,fill=gray!5},
    inout/.style={rounded corners=1pt,rectangle,draw=black!90,
        fill=violet!5,minimum width=0.6cm, minimum height=0.6cm},
    skip/.style={line width=0.2mm, ->},
]
    \def\yshift{0.3em}
    \def\base{8cm}
    \def\deltax{1cm}
    \def\dec{0.55cm}

    \node (base) at (0, 0) {};
    \node (pad) at (0cm, -1cm) {};
    \node (conv) [conv, minimum width=\base - \dec, anchor=south] at
      (base.north) {$\ReLU(\Conv(C_{in}, C_{out}, K=8, S=4))$};
    \node (rewrite) [rewrite, minimum width=\base - 2 * \dec, anchor=south] at
      ($(conv.north) + (0,\yshift)$) {$\GLU(\Conv(C_{out}, 2 C_{out}, K=1, S=1))$};

    \def\yshift{0.6em}
    \node (skip) [inout, anchor=south] at ($(rewrite.north) + (-\deltax, \yshift)$) {$\mathrm{Decoder}_i$};
    \draw[->] ($(rewrite.north) + (-\deltax, 0)$) -- (skip.south);

    \node (prev) [inout, anchor=north] at ($(conv.south) - (0, \yshift)$) {$\mathrm{Encoder}_{i-1}$ or input};
    \draw[->]  (prev.north) -- (conv.south);

    \node (next) [inout, anchor=south] at ($(rewrite.north) + (\deltax, \yshift)$) {$\mathrm{Encoder}_{i+1}$ or LSTM};
    \draw[->]  ($(rewrite.north) + (\deltax, 0)$) -- (next.south);
\end{tikzpicture}
    \caption{Detailed view of the layers $\mathrm{Decoder}_i$ on the top and
      $\mathrm{Encoder}_i$ on the bottom. Arrows represent connections
    to other parts of the model. For convolutions, 
    $C_in$ (resp $C_out$) is the number of input channels (resp output), $K$ the kernel size and $S$ the stride.\label{fig:modeldetail}}
  \end{subfigure}
  \caption{Demucs complete architecture on the left,
    with detailed representation of the encoder and decoder layers
  on the right.
}
\end{figure}
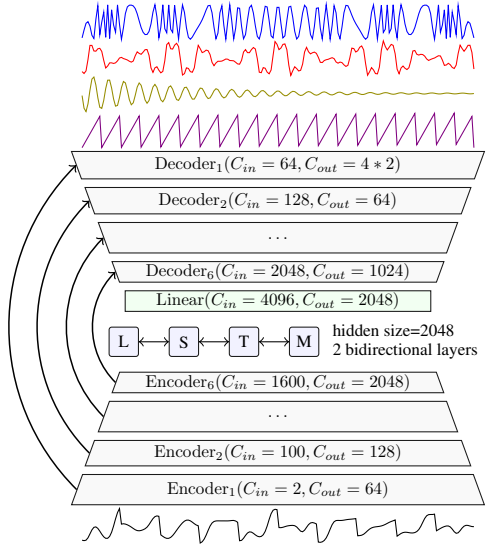
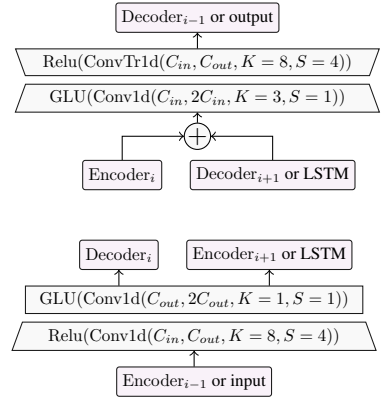

The architecture we propose, which we call Demucs, is described in the next few subsections, and the reconstruction loss is discussed in Section \ref{sec:loss}.
Demucs takes a stereo mixture as input
and outputs a stereo estimate for each source ($C=2$). It is an encoder/decoder architecture composed
of a convolutional encoder, a bidirectional LSTM, and a convolutional decoder, with
the encoder and decoder linked with skip U-Net connections. 
Similarly to other work in generation in both image~\citep{gan_style,progressive_gan} and sound~\citep{sing},
we do not use batch normalization~\citep{batchnorm} as our early experiments
showed that it was detrimental to the model performance. The overall architecture is depicted in Figure~\ref{fig:fullmodel}.

\subsection{Convolutional auto-encoder}
\label{sec:auto}

\paragraph{Encoder}
The encoder is composed of $L := 6$ stacked convolutional blocks numbered from 1 to $L$.
Each block $i$ is composed of
a convolution with kernel size $K:=8$,  stride $S:=4$, $C_{i-1}$ input channels,
$C_{i}$ output channels and ReLU activation, 
followed by a convolution with kernel size $1$, $2 C_i$ output channels and gated linear units (GLU) as activation function~\citep{glu}. Since GLUs halve the number of channels, the final output of block $i$ has $C_i$ output channels. A block is described in Figure~\ref{fig:modeldetail}.
Convolutions with kernel width 1 increase the depth and expressivity of the model at low computational cost. As we show in our ablation study \ref{sec:ablation}, the usage of GLU activations after these convolutions significantly boost performance. 
The number of channels in the input mixture is  $C_0 = C = 2$, while we use $C_1:=64$
as the number of output channels for the first encoder block. The number of channels is then doubled at each subsequent block, i.e.,  $C_i := 2 C_{i-1}$  for $i=2..L$, so the final number of channels is $C_L = 2048$.
We then use a bidirectional LSTM with 2 layers and a hidden size $C_L$.
The LSTM outputs $2 C_L$ channels per time position. We use a linear layer to take that number down to $C_L$.

\paragraph{Decoder}
The decoder is mostly the inverse of the encoder.
It is composed of $L$ blocks numbered in reverse order from $L$ to $1$.
The $i$-th blocks starts with a convolution with stride $1$ and kernel width $3$ to provide context about adjacent time steps,
input/output channels $C_i$ and a ReLU activation.
Finally, we use a transposed convolution
with kernel width $8$ and stride $4$, $C_{i-1}$ outputs and ReLU activation.
The $S$ sources are synthesized at the final layer only, after all decoder blocks. The final layer is linear with $S \cdot C_0$ output channels, one for each source ($4$ stereo channels in our case), without any additional activation function. Each of these channels directly generate the corresponding waveform.

\paragraph{U-network structure}

Similarly to Wave-U-Net \citep{waveunet_singing}, there are skip connections between the encoder and decoder blocks with the same index, as originally proposed in U-networks \citep{unet}. While the main motivation comes from empirical performances, an advantage of the skip connections is to give a direct access to the original signal, and in particular allows to directly transfers the phase of the input signal to the output, as discussed in Section~\ref{sec:loss}.
Unlike Wave-U-Net, we use transposed convolutions rather than linear interpolation followed by a convolution with a stride of 1. For the same increase in the receptive field, transposed convolutions require 4 times less operations and memory. This limits the overall number of channels that can be used
before running out of memory. As we observed that a large number of channels was key to obtaining good results, we favored the use of transposed convolutions, as explained in Section~\ref{sec:results}.

\paragraph{Motivation: synthesis vs masking}

The approach we follow uses the U-Network architecture \citep{unet,wavunet,waveunet_singing}, and builds on transposed convolutions with large number of channels and large strides ($4$) inspired by the approach to the synthesis of music notes of \citet{sing}. 
The U-Net skip connections and the gating performed by GLUs imply that this architecture is expressive enough to represent masks on a learnt representation of the input signal,
in a similar fashion to Conv-Tasnet. The Demucs approach is then more expressive than Conv-Tasnet, and its main advantages are the multi-scale representations of the input and the non-linear transformations to and from the waveform domain.

\paragraph{Resampling}
We observed improved performance when performing upsampling of the input by a factor of 2, and downsampling of the output to recover
the proper sample rate. We perform this operation as part of the end-to-end training loss, with a sinc resampling filter~\cite{smith1984flexible}. Note that the same
positive impact of resampling has been noted for speech denoising~\citep{defossez2020real}.

\subsection{Loss function}
\label{sec:loss}

For the reconstruction loss  $L(g_s(x; \theta), x_s)$ in \eqref{eq:separation}, we either use the average mean square error or average absolute error between waveforms: for a waveform $x_s$ containing $T$ samples and corresponding to source $s$, a predicted waveform $\hat{x}_s$ and denoting with a subscript $t$ the $t$-th sample of a waveform, we use one of $L_1$ or $L_2$:
\begin{equation}
    L_1(\hat{x}_s, x_s) = \frac{1}{T} \sum_{t=1}^T |\hat{x}_{s,t} - x_{s,t}|~~~~~~~~~~~~
    L_2(\hat{x}_s, x_s) = \frac{1}{T} \sum_{t=1}^T (\hat{x}_{s,t} - x_{s,t})^2\,.
\end{equation}

In generative models for audio, direct reconstruction losses on waveforms can pose difficulties because they are sensitive to the initial phases of the signals: two signals whose only difference is a shift in the initial phase are perceptually the same, but can have arbitrarily high $L_1$ or $L_2$ losses. It can be a problem in pure generation tasks because the initial phase of the signal is unknown, and losses on power/magnitude spectrograms are alternative that do not suffer from this lack of specification of the output. Approaches that follow this line either generate spectrograms \citep[e.g.,][]{tacotron}, or use a loss that compares power spectrograms of target/generated waveforms \citep{sing}.

The problem of invariance to a shift of phase is not as severe in source separation as it is in unconditional generation, because the model has access to the original phase of the signal. The pase can easily be recovered from the skip connections in U-net-style architectures for separation, and is directly used as input of the inverse STFT for methods that generate masks on power spectrograms. As such, losses such as $L1/L2$ are totally valid for source separation. Early experiments with an additional term including the loss of \citet{sing} did not suggest that it boosts performance, so we did not pursue this direction any further. Most our experiments use $L1$ loss, and the ablation study presented in Section \ref{sec:ablation} suggests that there is no significant difference between $L1$ and $L2$.

\subsection{Weight rescaling at initialization}
\label{sec:weight_init}

The initialization of deep neural networks is known to have a critical impact on the overall performances \citep{glorot2010understanding,kaiming_init}, up to the point that \citet{zhang2019fixup} showed that with a different initialization called fixup, very deep residual networks and transformers can be trained without batch normalization. While Fixup is not designed for U-Net-style skip connections, we observed that the following different initialisation scheme had great positive impact on performances compared to the standard initialization of \citet{kaiming_init} used in U-Networks. 

Considering the so-called Kaiming initialization \citep{kaiming_init} as a baseline, let us look at a single convolution layer for which we denote $w$ the weights after the first initialization. We take $\alpha := \mathrm{std}(w)/a$,
where $a$ is a reference scale, and replace $w$ by $w' = w / \sqrt{\alpha}$. Since the original weights have element-wise order of magnitude $(K C_{\rm in})^{-1/2}$ where $K$ is the kernel width and $C_{\rm in}$ the number of output channels, it means that our initialization scheme produces weights of order of magnitude $(K C_{\rm in})^{-1/4}$, together with a non-trivial scale. Based a search over the values [0.01, 0.05, 0.1], we select $a=0.1$ for all the regular
and transposed convolutions, see Section~\ref{sec:results} for more details. We experimentally observed that on a randomly initialized
model applied to an audio extract, it kept the standard deviation of the features along the layers of the same order of magnitude. Without initial rescaling,
the output the last layer has a magnitude 20 times smaller than the first.

\subsection{The shift trick}
\label{sec:equi}

A perfect source separation model is time equivariant, i.e. shifting the input mixture by X samples will shift the output Y by the exact same amount.
Thanks to  its dilated convolutions with a stride of 1, the mask predictor of Conv-Tasnet is naturally time equivariant and even if the encoder/decoder
is not strictly equivariant, Conv-Tasnet still verifies this property experimentally~\citep{convtasnet}. Spectrogram based method will also verify approximately this property. Shifting the input by a small amount will only reflect in the phase of the spectrogram. As the mask is computed only from the magnitude, and the input mixture phase is reused, the output will naturally be shifted by the same amount.
On the other hand, we noticed that our architecture did not naturally satisfy this property. We propose a simple workaround, the \emph{shift trick}\footnote{
Fabian-Robert Stöter is to be credited for the name.} , where we sample $S$ random shifts of an input mixture $x$ and average the predictions of our model for each, after having applied the opposite
shift. This technique does not require changing the training procedure or network architecture. Using $S=10$, we obtained a 0.3 SDR gain, see Section~\ref{sec:ablation} for more details. It does make evaluation of the model $S$ times slower, however, on a V100 GPU, separating
1 minute of audio at 44.1 kHz with Demucs takes only 0.8 second. With this technique, separation of 1 minute takes 8 seconds which is still more than 7 times faster
than real time.

\section{Experimental setup}
\label{sec:exp}

\subsection{Evaluation framework}

\paragraph{MusDB and additional data}
\label{sec:data}
We use the MusDB dataset~\citep{musdb} , which is composed of 150 songs
with full supervision in stereo and sampled at
44100Hz. For each song, we have the exact waveform of the \source{drums}, \source{bass}, \source{other} and \source{vocals} parts, i.e. each of the sources. The actual song, the mixture,
is the sum of those four parts. The first 84 songs form the \emph{train set}, the next 16 songs form the \emph{valid set} (the exact split is defined in the \texttt{musdb} python package) while the remaining 50 are
kept for the \emph{test set}.
We collected raw stems for 150 tracks, i.e., individual
instrument recordings used in music production software to make a song.
We manually assigned each instrument to one of the sources in MusDB.
We call this extra supervised data the \emph{stem set}.
We also report the performances of Tasnet and Demucs trained
using these 150 songs in addition to the 84 from MusDB, to anaylze the effect of adding more training data.

\paragraph{Source separation metrics}
\label{sec:metrics}
Measurements of the performance of source separation models was
developed by Vincent et al.~for blind source separation~\citep{measures}
and reused for supervised source separation in the SiSec Mus evaluation campaign~\citep{sisec}.
Similarly to previous work~\citep{wavunet,sony_densenet,sony_denselstm}, we focus on the SDR (Signal to Distortion Ratio) which measures the log ratio between the volume
of the estimated source projection onto the ground truth, and the volume of what is left out of this projection, typically contamination by other sources or artifacts. Other metrics can be defined ($\SIR$ and $\SAR$) and we present them in the supplementary material.
We used the python package
\texttt{museval}
 which provide a reference implementation for the SiSec Mus 2018 evaluation campaign.
As done in the SiSec Mus competition, we report the median over all tracks of the median
of the metric over each track computed using the \texttt{museval} package.

\subsection{Baselines}

We compare to several state-of-the-art baselines, both in the spectral and temporal domain.
Open Unmix~\citep{openunmix}, a 3-layer BiLSTM model with encoding and decoding fully connected layers
on spectrogram frames. It was released by the organizers of the SiSec 2018~\citep{sisec} to act as a strong reproducible baseline and matches the
performances of the best candidates trained only on MusDB.
We also selected MMDenseLSTM~\citep{sony_denselstm}, a multi-band dense net with LSTMs at different scales of the encoder and decoder.
This model was submitted as \textsc{TAK2} and trained with 804 extra labeled songs\footnote{Source: \url{https://sisec18.unmix.app/\#/methods/TAK2}}.

We also compare to the more recent options like the D3Net~\citep{takahashi2020d3net} model, which used dilated convolutions with dense connections, as well as the Spleeter model~\citep{spleeter2020}, a U-Net model trained on a very diverse dataset.
MMDenseLSTM, D3Net, Spleeter, and Open Unmix use Wiener filtering~\citep{multichannel_deep} as a last post processing step.
We provide the metrics for the Ideal Ratio Mask oracle (IRM), which computes the best possible mask using the ground truth sources
and is the topline of spectrogram based method~\citep{sisec}.

The only waveform based method submitted to the SiSec 2018 evaluation campaign is Wave-U-Net~\citep{wavunet} with the identifier \textsc{STL2}.
We also compare to the more recent variant of DPRNN by \citet{nachmani2020voice}, Meta-Tasnet~\citep{samuel2020meta}, and
Tasnet trained on extra data~\citep{lancaster2020frugal}.


\subsection{Training procedure}
\label{sec:training}

\paragraph{Epoch definition and augmentations}
One epoch over the dataset is defined as a pass over all 11-second extracts with a stride of 1 second. We use a random time shift between
0 and 1 second and keep 10 seconds of audio from there as a training example. We perform the following data augmentation~\citep{uhlich2017improving}, also used by Open Unmix and MMDenseLSTM: shuffling sources within one batch to generate a new mix, randomly swapping channels, random scaling by a uniform factor between 0.25 and 1.25. We additionally
multiply each source by $\pm 1$~\citep{nachmani2019unsupervised}. 
Following~\citep{cohen2019improving}, we also experiment with pitch/tempo shift. 20\% of the time, we randomly change the pitch by
-2, -1, 0, , +1, or +2 semitones, and the tempo by a factor taken uniformly between 0.88 and 1.12, using the Soundstretch utility\footnote{\url{https://www.surina.net/soundtouch/soundstretch.html}}.
When trained with extra data, we do not use pitch/tempo augmentation for Conv-Tasnet as it deteriorates its performance.

All Demucs models were trained over 360 epochs. Conv-Tasnet was trained for 360 epochs when trained only on MusDB and 240 when trained with extra data and using only 2-seconds audio extracts, due to its high memory usage.

\paragraph{Training setup and hyper-parameters}
The Demucs models are trained with 8 V100 GPUs with 16GB of RAM, while the Conv-Tasnet
models were trained with 16 V100 with 32GB of RAM. We use a batch size of 64, the Adam~\citep{adam} optimizer with a learning rate of 3e-4. 
For the Demucs model, we experimented with an initial number of channels in [32, 48, 64], with the best performance achieved for 64 channels.
Based on early experiments, we set the initial weight rescaling reference level described in Section~\ref{sec:weight_init}, to 0.1.
We computed confidence intervals using 3 random seeds in Table~\ref{table:comparison}.
For the ablation study on Table~\ref{table:ablation}, we provide metrics for a single run. 

\paragraph{Quantization}

For quantization, we use the DiffQ method~\citep{defossez2021differentiable}. DiffQ uses additive
noise on the weights as a proxy for the true quantization error.
The scale of the noise depends on learnable parameters that controls the number of bits used per group of 8 weight entries.
Because this transformation is completely differentiable, one can simply use a model size penalty (expressed in MegaBytes), with a weight of 0.0003, added to the main reconstruction loss.

\section{Experimental results}
\label{sec:results}

\begin{table}
\caption{Comparison of Conv-Tasnet and Demucs to state-of-the-art models that operate on the waveform (Wave-U-Net, Meta-Tasnet, DPRNN-like, Tasnet trained with extra data) and on spectrograms (Open-Unmix without extra data, D3Net, MMDenseLSTM with extra data) as well as the IRM oracle on the MusDB test set.
The \emph{Extra?} indicates the number of extra training songs used. We report the median over all tracks
of the median $\SDR$ over each track, as done in the SiSec Mus evaluation campaign~\citep{sisec}.
The \source{All} column reports the average over all sources. 
Demucs metrics are averaged over 3 runs, the confidence interval is the standard deviation over $\sqrt{3}$.
In bold are the values that are statistically state-of-the-art either with or without extra training data.}
\label{table:comparison}
\begin{center}
\resizebox{0.98\textwidth}{!}{
\begin{tabular}{l c c l l l l l}
  \toprule
     &&& \multicolumn{5}{c}{Test $\SDR$ in dB}\\
     \cmidrule{4-8}
  \textbf{Architecture} & \textbf{Wav?} & \textbf{Extra?} &
  \source{All} & \source{Drums} &  \source{Bass} &\source{Other} & \source{Vocals}\\
  \midrule
  IRM oracle& \crmark & N/A & 8.22 & 8.45 & 7.12 & 7.85 & 9.43\\
  \midrule
  Wave-U-Net & \chmark & \crmark & 3.23 & 4.22 & 3.21 & 2.25 & 3.25\\
  Open-Unmix & \crmark & \crmark & 5.33 & 5.73 & 5.23 & 4.02 & 6.32\\
  Meta-Tasnet & \chmark & \crmark & 5.52 & 5.91 & 5.58 &  4.19 & 6.40\\
  Conv-Tasnet$^\dagger$ & \chmark & \crmark & 5.73 $\scriptstyle\pm .10$ & 6.02 $\scriptstyle\pm .08$ & 6.20 $\scriptstyle\pm .15$ & 4.27 $\scriptstyle\pm .03$ & 6.43 $\scriptstyle\pm .16$ \\
  DPRNN & \chmark & \crmark & 5.82 &6.15 & 5.88 & 4.32 & 6.92 \\
  D3Net & \crmark & \crmark & 6.01 &\textbf{7.01} & 5.25 & \textbf{4.53} & \textbf{7.24} \\
  Demucs$^\dagger$ & \chmark & \crmark & 6.28 $\scriptstyle\pm .03$ & 6.86 $\scriptstyle\pm .05$ & \textbf{7.01} $\scriptstyle\pm .19$ & 4.42 $\scriptstyle\pm .06$ & 6.84 $\scriptstyle\pm .10$ \\
  \midrule
  Spleeter & \crmark & $\sim 25\mathrm{k}^*$ & 5.91 & 6.71 & 5.51 & 4.55 & 6.86\\
  TasNet & \chmark & $\sim 2.5\mathrm{k}$ & 6.01 & 7.01 & 5.25 & 4.53 & 7.24\\
  MMDenseLSTM & \crmark & 804 & 6.04 & 6.81 & 5.40 & 4.80 & 7.16\\
  Conv-Tasnet$^{\dagger\dagger}$ & \chmark & 150 & 6.32 $\scriptstyle \pm .04$ & 7.11 $\scriptstyle\pm .13$ & 7.00 $\scriptstyle\pm .05$ & 4.44$\scriptstyle\pm .03$ & 6.74 $\scriptstyle\pm .06$ \\
  D3Net & \crmark & 1.5k & 6.68 & 7.36  & 6.20 & \textbf{5.37} & \textbf{7.80} \\
  Demucs$^{\dagger}$ & \chmark & 150 & \textbf{6.79} $\scriptstyle\pm .02$ & \textbf{7.58} $\scriptstyle\pm .02$ & \textbf{7.60} $\scriptstyle\pm .13$ & 4.69 $\scriptstyle\pm .04$ & 7.29 $\scriptstyle\pm .06$ \\
  \bottomrule
\end{tabular}}
\end{center}
*: each track is only 30 seconds,
$\dagger$: from current work, $\dagger\dagger$: trained without pitch/tempo augmentation, as it deteriorates performance.
\end{table}

\begin{table}
\caption{Mean Opinion Scores (MOS) evaluating the quality and absence of artifacts of the separated audio. 38 people rated 20 samples each, 
randomly sample from one of the 3 models or the ground truth. There is one sample per track in the MusDB
test set and each is 8 seconds long. Ratings of 5 means that the quality is perfect (no artifacts).}
\label{table:mos_quality}
\begin{center}
\begin{tabular}{l l l l l l}
  \toprule
     & \multicolumn{5}{c}{Quality Mean Opinion Score}\\
     \cmidrule{2-6}
  \textbf{Architecture} & 
  \source{All} & \source{Drums} &  \source{Bass} &\source{Other} & \source{Vocals}\\
  \midrule
  Ground truth & 4.46 $\scriptstyle\pm .07$ & 4.56 $\scriptstyle\pm .13$ & 4.25 $\scriptstyle\pm .15$ & 4.45 $\scriptstyle\pm .13$ & 4.64 $\scriptstyle\pm .13$ \\
  \midrule
  Open-Unmix & 3.03 $\scriptstyle\pm .09$ & 3.10 $\scriptstyle\pm.17$ & 2.93 $\scriptstyle\pm .20$ & 3.09 $\scriptstyle\pm 0.16$ & 3.00 $\scriptstyle\pm .17$ \\
  Demucs & 3.22 $\scriptstyle\pm .09$ & 3.77 $\scriptstyle\pm.15$ & 3.26 $\scriptstyle\pm .18$ & 3.32 $\scriptstyle\pm .15$ &  2.55 $\scriptstyle\pm .20$ \\
 Conv-Tasnet & 2.85 $\scriptstyle\pm.08$ & 3.39 $\scriptstyle\pm.14$ & 2.29 $\scriptstyle\pm .15$ & 3.18 $\scriptstyle\pm .14$ & 2.45 $\scriptstyle\pm .16$\\
  \bottomrule
\end{tabular}
\end{center}
\end{table}
\begin{table}
\caption{Mean Opinion Scores (MOS) evaluating contamination by other sources. 38 people rated 20 samples each, 
randomly sampled from one of the 3 models or the ground truth. There is one sample per track in the MusDB
test set and each is 8 seconds long. Ratings of 5 means no contamination by other sources.}
\label{table:mos_contamination}
\begin{center}
\begin{tabular}{l l l l l l}
  \toprule
     & \multicolumn{5}{c}{Contamination Mean Opinion Score}\\
     \cmidrule{2-6}
  \textbf{Architecture} & 
  \source{All} & \source{Drums} &  \source{Bass} &\source{Other} & \source{Vocals}\\
  \midrule
  Ground truth & 4.59 $\scriptstyle\pm .07$ & 4.44 $\scriptstyle\pm .18$ & 4.69 $\scriptstyle\pm .09$ & 4.46 $\scriptstyle\pm .13$ & 4.81 $\scriptstyle\pm .11$ \\
  \midrule
  Open-Unmix & 3.27 $\scriptstyle\pm .11$ & 3.02 $\scriptstyle\pm.19$ & 4.00 $\scriptstyle\pm .20$ & 3.11 $\scriptstyle\pm .21$ & 2.91 $\scriptstyle\pm .20$ \\
  Demucs & 3.30 $\scriptstyle\pm .10$ & 3.08 $\scriptstyle\pm.21$ & 3.93 $\scriptstyle\pm .18$ & 3.15 $\scriptstyle\pm .19$ &  3.02 $\scriptstyle\pm .20$ \\
 Conv-Tasnet & 3.42 $\scriptstyle\pm.09$ & 3.37 $\scriptstyle\pm.17$ & 3.73 $\scriptstyle\pm .18$ & 3.46 $\scriptstyle\pm .17$ & 3.10 $\scriptstyle\pm .17$\\
  \bottomrule
\end{tabular}
\end{center}
\end{table}

In this section, we first provide experimental results on the MusDB dataset for Conv-Tasnet and Demucs compared with state-of-the-art baselines. We then dive into the ablation study of Demucs. 

\subsection{Comparison with baselines}

We provide a comparison with the state-of-the-art baselines on Table~\ref{table:comparison}. The models on the top half were trained without any extra data
while the lower half used unreleased training songs. As no previous work included confidence intervals, we considered the single metric provided by for the baselines as the exact estimate of their mean performance.

\paragraph{Quality of the separation}

We first observe that Demucs and Conv-Tasnet outperform previous waveform domain methods such as Wave-U-Net. Demucs surpasses the best spectrogram domain method, D3Net~\citep{takahashi2020d3net}, when averaging over all sources, by 0.3 dB of SDR.
When looking into more details, we see a clear advantage for spectrogram domain methods for the \source{other} and \source{vocals} sources. Without extra training data, D3Net also beats Demucs for the \source{drums}. However, we will see in the next section that human evaluations show that spectrogram domain methods tend to deteriorate the attack of an instrument, so that it is still possible that this high SDR hides significant degradation of the attack of percussive sounds. As no audio samples were released for D3Net, this is impossible to verify.

When trained with extra training data, we still see a small advantage overall from Demucs over D3Net (0.1 point of SDR). Again we notice that the \source{bass} and \source{drums} sources are better handled in the temporal domain, while \source{other} and \source{vocals} are better handled in the spectrogram domain.
Note that for the first time for music source separation, a temporal domain model beats the IRM oracle for the \source{bass} source, by 0.5 points of SDR.

An interesting experiment would be retrain D3Net with pitch/tempo augmentation and see if it benefits from it. Interestingly, the Conv-Tasnet model achieved the same SDR whether this augmentation was used or not, and the SDR even deteriorated when trained with extra tracks. For Demucs, which comparatively has more tunable weights (see Table~\ref{table:quantization}), the augmentation always improves performance, most likely by preventing some overfitting. We conjecture that pitch/tempo shift lead to deteriorated audio, but reduce overfitting. For Demucs, this is a clear win, while for smaller models, it might not be as beneficial.

\paragraph{Human evaluations}

We noticed strong artifacts on the audio separated by Conv-Tasnet, especially for the \source{drums} and \source{bass} sources: static noise between 1 and 2 kHz, 
hollow instrument attacks or missing notes as illustrated on Figure~\ref{fig:mel}. In order to confirm this observation, we organized a mean opinion score survey. 
We separated 8 seconds extracts from all of the 50 test set tracks for Conv-Tasnet, Demucs and Open-Unmix. We asked 38 participants to rate 20 samples each, randomly taken from one
of the 3 models or the ground truth. For each one, they were required to provide 2 ratings on a scale of 1 to 5. The first one evaluated the quality and absence of artifacts (1: many artifacts and distortion, content is hardly recognizable, 5: perfect quality, no artifacts) and the second one evaluated contamination by other sources (1: contamination if frequent and loud, 5: no contamination). We show the results on Tables~\ref{table:mos_quality} and \ref{table:mos_contamination}.

Note that for this experiment, with used an early version of Demucs that was trained with 100 channels, and without pitch/tempo augmentation. We had trouble differentiating between the latest version of Demucs and this early version in an informal blind test performed by the authors, so that we believe the results presented here stay mostly valid for the best performing Demucs model presented on Table~\ref{table:comparison}.

We confirmed that the presence of artifacts in the output of Conv-Tasnet degrades the user experience, with a rating of 2.85$\pm .08$ against $3.22\pm.09$ for Demucs.
On the other hand, Conv-Tasnet samples had less contamination by other sources than Open-Unmix or Demucs, although by a small margin, with a rating of $3.42\pm.09$
against $3.30\pm.10$ for Demucs and $3.27\pm.11$ for Open-Unmix.

\paragraph{Training speed} We measured the time taken to process a single batch of size 16 with 10 seconds of audio at 44.1kHz (the original Wave-U-Net being only trained on 22 kHz audio, we double the time for fairness), on a single GPU, ignoring data loading and using \verb|torch.cuda.synchronize|
to wait on all kernels to be completed. MMDenseLSTM does not provide a reference implementation. Wave-U-Net takes 1.2 seconds per batch, Open Unmix 0.2 seconds per batch
and Demucs 1.4 seconds per batch. Conv-Tasnet cannot be trained with such a large sample size, however a single iteration over 2 seconds of audio with a batch size of 4
takes 0.7 seconds.

\subsection{Ablation study for Demucs}
\label{sec:ablation}
\begin{table}
\caption{Ablation study for the novel elements in our architecture described in Section~\ref{sec:model}.
We use only the train set from MusDB and report best L1 loss over the valid set throughout training
as well the SDR on the test set for the epoch that achieved this loss.
}
\label{table:ablation}
\begin{center}
  \setlength\extrarowheight{1pt}
\begin{tabular}{l r r}
  \toprule
  & \textbf{Valid set} & \textbf{Test set}\\
  \textbf{Difference}& L1 loss  & $\SDR$\\
  \midrule
  no BiLSTM & 0.171 & 5.40\\
  no pitch/tempo aug. & 0.156 & 5.80 \\
  no initial weight rescaling & 0.156 & 5.88\\
  no 1x1 convolutions in encoder &  0.154 & 5.89 \\
  ReLU instead of GLU &  0.157 & 5.92 \\
  no BiLSTM, depth=7  & 0.160 & 5.94\\
  MSE loss  & N/A & 5.99\\
  no resampling & 0.153 & 6.03 \\
  no shift trick &  N/A & 6.05 \\
  kernel size of 1 in decoder convolutions & 0.153 & 6.11 \\
  \midrule
  Reference & 0.150 & 6.28\\
  \bottomrule
\end{tabular}
\end{center}
\vskip -3mm
\end{table}

\begin{table}
\caption{
Impact of the initial number of channels on the model size, and on the performance. We also report
the results obtained from quantizing the model with DiffQ.
The bottom part consist in the base Demucs and Conv-Tasnet models presented on Table~\ref{table:comparison}.
}
\label{table:quantization}
\begin{center}
  \setlength\extrarowheight{1pt}
\begin{tabular}{l r r r}
  \toprule
  & & \textbf{Valid set} & \textbf{Test set}\\
  \textbf{Model}& \textbf{Model size (MB)} & L1 loss  & $\SDR$\\
  \midrule
  Demucs, 32 channels & 254 & 0.154 & 6.08 \\
  Demucs, 48 channels &  570 & 0.151 & 6.20\\
  \midrule
  Demucs, 64 channels, DiffQ quantized & 120 & \textbf{0.150} & \textbf{6.28} \\ 
  \midrule
  Demucs, 64 channels & 1014 & \textbf{0.150} & \textbf{6.28}\\
  Conv-Tasnet & 42 & 0.152 & 5.73\\
  \bottomrule
\end{tabular}
\end{center}
\vskip -3mm
\end{table}

We provide an ablation study of the main design decisions for Demucs in Table~\ref{table:ablation}. Given the cost of training a single model, we did not
compute confidence intervals for each variation.

We notice strong contributions for different design decisions we made. The L1 loss seems to be more robust than the MSE. The LSTM part plays a crucial role,
and replacing it by an extra layer of convolution is not sufficient to recover the best possible SDR.
The initial weight rescaling described in Section~\ref{sec:weight_init} provides significant gain, with an extra 0.4 dB for the SDR.

We introduced extra convolutions in the encoder and decoder, as described in Sections~\ref{sec:auto}. The two proved useful, 
improving the expressivity of the model, especially when combined with GLU activation.
Using a kernel size of 3 instead of 1 in the decoder
further improves performance. We conjecture that the context from adjacent time steps helps the output of the transposed convolutions to be consistent through time
and reduces potential artifacts arising from using a stride of 4.

The pitch/tempo augmentation gives one of the largest gain, with almost 0.5 dB extra for the SDR, which highlights the importance of strong data augmentation when training data is limited.

We observe that using the shift trick as described in Section~\ref{sec:model} gives a gain of almost 0.3 SDR.
We did not report the validation loss as we only use the stabilization when computing the SDR over the test set.
We applied the randomized stabilization to Open-Unmix and Conv-Tasnet with no gain, since, as explained in Section~\ref{sec:equi}, both are naturally equivariant with respect to initial time shifts.



\subsection{Quantization of Demucs}

One of the main drawbacks of the Demucs model when compared to other architecture is its large model size, more than 1014MB, against 42MB for Conv-TasNet. On Table~\ref{table:quantization}, we propose 
two ways to reduce this issue, either by reducing the initial number of channels (32 or 48 channels), which will improve both the model size, as well as reduce the computational complexity of the model, or using the DiffQ quantization technique~\citep{defossez2021differentiable}.

While it is possible to use 48 channels without suffering from a large loss in SDR, it seems getting to 32 channels will lead to a decrease of 0.2 dB in performance.
On the other hand, quantization will reduce the model size down to 120MB without any loss of SDR. This is still more than the 42MB of Conv-Tasnet, but close to 10x improvement over the uncompressed baseline.

Note that the 120MB is achieved if each integer is exactly
coded on the right number of bits. As this is currently
not implemented in Python, we rely on \texttt{gzip} to
approximately reach this ideal size. This leads to a small overhead, so that the download size for our quantized model is
150MB.

\section*{Conclusion}

We experimented with two architectures for music source separation in the waveform domain: Demucs and Conv-Tasnet. We show that with proper data augmentation, Demucs surpasses all state-of-the-art architecture in the waveform or spectrogram domain by at least 0.3 dB of SDR. However, their is no clear winner between waveform and spectrogram domain models, as the former seems to dominate for the \source{bass} and \source{drums} sources, while the latter obtain the best performance on the \source{vocals} and \source{other} sources, as measured both by objective metrics and human evaluations. We conjecture that spectrogram domain models have an advantage when the content is mostly harmonic and fast changing, while for sources without harmonicity (drums) or with strong and emphasized attack regimes (bass), waveform domain will better preserve the structure of the music source.

In terms of training and architecture, we confirm the importance of using pitch/tempo shift augmentations (although the Conv-Tasnet architecture does not seem to benefit from it), as well as using LSTM for long range dependencies, and powerful encoding and decoding layers with 1x1 convolutions and GLU activations.

When trained with extra data, Demucs surpasses for the first time the IRM oracle for the bass source. On the other hand, Demucs still suffers from larger leakage than other architectures,
specifically for the \source{vocals} and \source{other} sources, which we will try to reduce in future work.

\clearpage
\bibliography{references}

\clearpage
\renewcommand{\thesection}{\Alph{section}}

\section*{\Large Appendix}
\setcounter{section}{0}
\renewcommand\theHsection{\Alph{section}}

\section{Results for all metrics}

Reusing the notations from ~\citet{measures},
let us take a source $j \in {1, 2, 3, 4}$ and
introduce $P_{s_j}$ (resp $P_\mathbf{s}$)
the orthogonal projection on ${s_j}$ (resp on $\mathrm{Span}(s_1, \ldots, s_4)$).
We then take with $\hat{s}_j$ the estimate of source~$s_j$
\begin{align*}
  s_{\mathrm{target}} := P_{s_j}(\hat{s}_j) \qquad
  e_{\mathrm{interf}} := P_{\mathbf{s}}(\hat{s}_j) - P_{s_j}(\hat{s}_j) \qquad
  e_{\mathrm{artif}} := \hat{s}_j - P_{\mathbf{s}}(\hat{s}_j)
\end{align*}
We can now define various signal to noise ratio, expressed in decibels (dB):
the source to distortion ratio
\[
  \SDR := 10 \log_{10}\frac{
      \norm{s_{\mathrm{target}}}^2}{
      \norm{e_{\mathrm{interf}} + e_{\mathrm{artif}}}^2},
\]
the source to interference ratio
\[
  \SIR := 10 \log_{10}\frac{
      \norm{s_{\mathrm{target}}}^2}{
      \norm{e_{\mathrm{interf}}}^2}
\]
and the sources to artifacts ratio
\[
  \SAR := 10 \log_{10}\frac{
    \norm{s_{\mathrm{target}} + e_{\mathrm{interf}}}^2}{
      \norm{e_{\mathrm{artif}}}^2}.
\]

As explained in the main paper, extra invariants are added when using the \texttt{museval} package. We refer
the reader to~\citet{measures} for more details.
In the following, we only provide the metrics for some of the baselines. We refer the reader to the original papers for the results for the other baselines.

\begin{center}
\begin{tabular}{l c c l l l l l}
  \toprule
     &&& \multicolumn{5}{c}{Test $\mathrm{SIR}$ in dB}\\
     \cmidrule{4-8}
  \textbf{Architecture} & \textbf{Wav?} & \textbf{Extra?} &
  \source{All} & \source{Drums} &  \source{Bass} &\source{Other} & \source{Vocals}\\
  \midrule
  IRM oracle & \crmark & N/A & 15.53 & 15.61 & 12.88 & 12.84 & 20.78\\
  \midrule
  Open-Unmix & \crmark & \crmark & 10.49 & 11.12 & 10.93 & 6.59 & 13.33\\
  Wave-U-Net & \chmark & \crmark & 6.26 & 8.83 & 5.78 & 2.37 & 8.06\\
  Demucs  & \chmark & \crmark & 11.93 $\scriptstyle\pm .04$ & 13.46 $\scriptstyle\pm .12$ & 13.42 $\scriptstyle\pm .16$ & 6.37 $\scriptstyle\pm .08$ & 14.48 $\scriptstyle\pm .07$ \\
  Conv-Tasnet & \chmark & \crmark & 11.75 $\scriptstyle\pm .20$ & 12.94 $\scriptstyle\pm .89$ & 12.62 $\scriptstyle\pm .19$ & 7.15 $\scriptstyle\pm .23$ & 14.27 $\scriptstyle\pm .59$ \\
  \midrule
  Demucs & \chmark & 150 & 12.86 $\scriptstyle\pm .05$ & 14.72 $\scriptstyle\pm .34$ & 14.36 $\scriptstyle\pm .08$ & 7.82 $\scriptstyle\pm .06$ & 14.53 $\scriptstyle\pm .22$ \\
  Conv-Tasnet & \chmark & 150 & 12.24 $\scriptstyle\pm .09$ & 13.66 $\scriptstyle\pm .14$ & 13.18 $\scriptstyle\pm .13$ & 8.40 $\scriptstyle\pm .08$ & 13.70 $\scriptstyle\pm .22$ \\
  MMDenseLSTM & \crmark & 804 & 12.24 & 11.94 & 11.59 & 8.94 & 16.48\\
  \bottomrule
\end{tabular}
\end{center}

\begin{center}
\begin{tabular}{l c c l l l l l}
  \toprule
     &&& \multicolumn{5}{c}{Test $\mathrm{SAR}$ in dB}\\
     \cmidrule{4-8}
  \textbf{Architecture} & \textbf{Wav?} & \textbf{Extra?} &
  \source{All} & \source{Drums} &  \source{Bass} &\source{Other} & \source{Vocals}\\
  \midrule
  IRM oracle & \crmark & N/A & 8.31 & 8.40 & 7.40 & 7.93 & 9.51 \\
  \midrule
  Open-Unmix & \crmark & \crmark & 5.90 & 6.02 & 6.34 & 4.74 & 6.52\\
  Wave-U-Net & \chmark & \crmark & 4.49 & 5.29 & 4.64 & 3.99 & 4.05\\
  Demucs & \chmark & \crmark & 6.08 $\scriptstyle\pm .01$ & 6.18 $\scriptstyle\pm .03$ & 6.41 $\scriptstyle\pm .05$ & 5.18 $\scriptstyle\pm .06$ & 6.54 $\scriptstyle\pm .04$ \\
  Conv-Tasnet & \chmark & \crmark & 6.13 $\scriptstyle\pm .04$ & 6.19 $\scriptstyle\pm .05$ & 6.60 $\scriptstyle\pm .07$ & 4.88 $\scriptstyle\pm .02$ & 6.87 $\scriptstyle\pm .05$ \\
  \midrule
  Demucs & \chmark & 150 & 6.50 $\scriptstyle\pm .02$ & 7.04 $\scriptstyle\pm .07$ & 6.68 $\scriptstyle\pm .04$ & 5.26 $\scriptstyle\pm .03$ & 7.00 $\scriptstyle\pm .05$ \\
  Conv-Tasnet & \chmark & 150 & 6.57 $\scriptstyle\pm .02$ & 7.35 $\scriptstyle\pm .05$ & 6.96 $\scriptstyle\pm .08$ & 4.76 $\scriptstyle\pm .05$ & 7.20 $\scriptstyle\pm .05$ \\
  MMDenseLSTM & \crmark & 804 & 6.50 & 6.96 & 6.00 & 5.55 & 7.48\\
  \bottomrule
\end{tabular}
\end{center}

\end{document}